\documentclass[]{pasj02} 
\draft
\usepackage[switch,mathlines]{lineno} % add line number to manuscript

\jyear{2024}
\Received{}%{yyyy/mm/dd}
\Accepted{}%{yyyy/mm/dd}
\begin{document} 

\title{Nuclear Stellar Disk-like Nature in the Kinematics of SiO Maser Stars around Sagittarius A$^\ast$ }

%%% begin:list of authors
% Do NOT capitalize all letters in "textsc".
\author{
Masato \textsc{Tsuboi},\altaffilmark{1}\altemailmark\orcid{0000-0001-8185-8954} \email{masato.tsuboi@meisei-u.ac.jp} 
Takahiro \textsc{Tsutsumi},\altaffilmark{2}\altemailmark\orcid{0000-0002-4298-4461}\email{ttsutsum@nrao.edu} 
Ryosuke \textsc{Miyawaki},\altaffilmark{3}\altemailmark\orcid{0000-0001-5259-4080}\email{miyawaki@obirin.ac.jp}
and
Makoto \textsc{Miyoshi}\altaffilmark{4}\altemailmark\orcid{0000-0002-6272-507X}\email{makoto.miyoshi@nao.ac.jp}
%and
%Atsushi \textsc{Miyazaki},\altaffilmark{5}\altemailmark\email{atsmiyazaki2@a-net.email.ne.jp}
}
%\footnotetext[$\dag$]{Present address: ....}
\altaffiltext{1}{School of Science and Engineering, Meisei University, 2-1-1 Hodokubo, Hino, Tokyo 191-8506, Japan}
\altaffiltext{2}{National Radio Astronomy Observatory (NRAO), P. O. Box O,  Socorro, NM 87801-0387, USA}
\altaffiltext{3}{College of Arts and Sciences, J. F. Oberlin University, Machida, Tokyo 194-0294, Japan}
\altaffiltext{4}{National Astronomical Observatory of Japan (NAOJ), Mitaka, Tokyo 181-8588, Japan}
%\altaffiltext{5}{Japan Space Forum, Kanda-surugadai, Chiyoda-ku, Tokyo,101-0062, Japan}
%%% end:list of authors

%% !!! Select 3 to 5 words from PASJ's key words !!! 
%% List of Key Words: https://academic.oup.com/pasj/pages/Pasj_Keywords 
%% "\KeyWords{ }" always has to be placed before ``\maketitle''
\KeyWords{Galaxy: center, Galaxy: kinematics and dynamics, stars: late-type, radio lines: stars}

\maketitle

\begin{abstract}
We present a detailed analysis of the kinematics of SiO maser stars around the center of the Milky Way, Sagittarius A$^\ast$ (Sgr A$^\ast$). We used the archive data in the SiO $v=1, J=2-1$ emission line obtained by the Atacama Large Millimeter/Submillimeter Array (ALMA) in 2017 and 2021 (\#2016.1.00940.S, PI Darling, J. and \#2019.1.00292.S, PI Paine, J.).  
We detected 37 SiO maser stars in the channel maps and derived their angular offsets relative to Sgr A$^\ast$ and LSR radial velocities. We derived the proper motions of 35 stars by comparing their angular offsets in the two epochs. 
The proper motions of Wolf-Rayet and O star in the Nuclear Star Cluster (NSC)  are reported to be rather random, except for the co-moving clusters IRS13E and IRS13N (Tsuboi et al. 2022). However, the derived proper motions of SiO maser stars do not look completely random. The proper motions of the SiO maser stars show a tendency to lie along the Galactic plane. The proper motion amplitudes of SiO maser stars are  larger than the LSR velocity amplitudes. 
We estimated the 3D motions from the proper motions and LSR velocities. Many 3D velocities are near to or larger than the upper limit velocities for Kepler orbits around Sgr A$^\ast$, whose mass is assumed to be $4\times10^6 M_\odot$.  These indicate that the SiO maser stars around Sgr A$^\ast$ are members of the Nuclear Stellar Disk rather than the NSC. 
\end{abstract}
%\pagewiselinenumbers 

\section{Introduction}
The Galactic Center (GC) region is the central region of the Milky Way galaxy, which is located at a distance of $d\sim8.2$ kpc from us (e.g., \cite{Boehle}, \cite{Abuter}). 
It harbors the Galactic Center Black Hole (GCBH), the super massive black hole with $M_{\mathrm{GCBH}}\sim4\times10^6 M_\odot$ (e.g. \cite{Ghez}, \cite{Boehle}, \cite{Abuter}). 
The counterpart of the GCBH is observed as Sagittarius A$^\ast$ (Sgr A$^\ast$) in radio, infrared (IR), and X-ray wavelengths.
The Atacama Large Millimeter/Submillimeter Array (ALMA) can image the region with the unprecedented small synthesized beam size. High-resolution observations of the GC region are important not only to study individual peculiar structures and phenomena in the region, but also to provide initial information for the study of central regions in spiral galaxies. 

There are stratified groups of stars in the GC region. The innermost group is called "S-stars" and is identified as a cluster of B stars within a few tenths of a pc of Sgr A$^\ast$ (e.g. \cite{Genzel1997}, \cite{Ghez1998}, \cite{Ghez2003}).  On the other hand, a cluster within a few pc of Sgr A$^\ast$ is called the Nuclear Star Cluster (NSC), which contains many O stars and Wolf-Rayet (WR) stars (e.g. \cite{Genzel1996}, \cite{Ghez1998}, \cite{Paumard}, \cite{Schodel2009}, \cite{Genzel2010}, \cite{Eckart2013}, \cite{Yelda2014}). Because it is difficult to form these stars in the strong tidal force environment of the GCBH, their origin is still controversial.  A disk-shaped group of stars is observed up to $l\sim\pm\timeform{1D}$ in the outer region (\cite{Launhardt}, \cite{Stolovy2006}, \cite{Nishiyama2013}). This is called the Nuclear Stellar Disk (NSD) and is composed mainly of older stars. These groups have been identified mainly on the basis of IR observations. There is an ongoing astrometric observing program with an IR space-borne telescope called JASMINE to determine the origin of the NSD (\cite{Gouda}, \cite{Kawata}).

Attempts to search for SiO maser stars in the region within a few pc of Sgr A$^\ast$ have been made with ATCA, JVLA and ALMA (e.g. \cite{Li2011}, \cite{Paine2022}, \cite{Darling2023}, \cite{Paine2024}).  The 3D kinematics of these stars can be determined from the radial velocity and the proper motion. The radial velocity of the maser star is easily observable, but that of the early-type stars is not easy to infer. 
ALMA has a very sharp synthesized beam with very high sensitivity. ALMA has a large field of view (FOV), $\timeform{20.6"}\times\frac{300}{\nu[\mathrm{GHz}]}=\timeform{67.5"}$ in Full Width Half Maximum (FWHM) for the SiO $v=1, J=2-1$ maser emission line ($\nu=86.2$ GHz) (see ALMA technical handbook), which corresponds to the area within the radius of $r\sim1.4$ pc of Sgr A$^\ast$. Moreover, the position of the stars can be measured directly by referencing the position of Sgr A$^\ast$ with ALMA, since ALMA can always detect Sgr A$^\ast$.   Therefore, ALMA is a suitable tool for precision astrometry within a few arcminutes of Sgr A$^\ast$ (e.g. \cite{Tsuboi2022}, \cite{Paine2022}, \cite{Darling2023}, \cite{Paine2024}). 

 Since the kinematics of these groups of stars are thought to be closely related to their origin, precision astrometry is important. The 3D kinematics of the maser stars in the vicinity of Sgr A$^\ast$ are very important for studying the mass distribution, including the enclosed mass in the region (e.g. \cite{Paine2022}, \cite{Darling2023}). However, these studies implicitly assume that these maser stars are members of the NSC only because they are observed in the direction of the NSC. The ages of the maser stars are expected to be $10^9$ years. On the other hand, the ages of the early type stars of the NSC are at most $10^8$ years. If these clusters of different ages belong to the same cluster, there is a mechanism for the mixing of early type stars and maser stars. Alternatively, the mixing may occur by chance only in the line of sight.
Here we focus on resolving this question based on statistical properties in the kinematics of SiO maser stars.
Throughout this paper, we adopt $d=8.2$ kpc as the distance to the Galactic center. $1\arcsec$ corresponds to $0.03975$ pc or  $1.224\times10^{17}$ cm at the distance. And we use the International Celestial Reference System (ICRS) as the coordinate system. 
 
 \section{Data acquisition and reduction}
 We have obtained the archive data in the SiO $J=2-1$ emission lines observed by the  ALMA in 2017 and 2021 (\#2016.1. 00940.S, PI Darling, J. and \#2019.1.00292.S, PI Paine, J.) from the ALMA Science Archive (https://almascience.nao.ac.jp/aq/).
 The ALMA observations consist of a single pointing toward Sgr A$^\ast$ with the 12-meter Array  in the SiO $v=0, J=2-1$ ($86.847010$ GHz), SiO $v=1, J=2-1$ ($86.243440$ GHz),  and SiO $v=2, J=2-1$ ($85.640446$ GHz)  emission lines. In addition, we  simultaneously observed the continuum emission of this region. We used the continuum map to derive the the position of Sgr A$^\ast$.
 The observation epochs in 2017 and 2021 were  September 19, 2017 and  August 19, 2021, respectively.
The original widths  of the velocity channel are $0.42$ km s$^{-1}$ for the $v=0$ and $v=2$ emission lines and $0.21$ km s$^{-1}$ for the SiO $v=1$ emission line, respectively.
J1733-1304,  J1744-3116, and J1924-2914 were used as a flux calibrator, a phase calibrator, and a band pass  calibrator, respectively.  

The calibration and imaging of the data were performed by CASA (\cite{CASA Team}). Before imaging in the the SiO maser lines, the continuum emission of Sgr A$^\ast$ and the surrounding continuum features were subtracted from the original data using the CASA task, {\tt uvcontsub (fitorder=0)}.
The beam sizes in FWHM using Briggs weighting (robust parameter = 0.5)  as {\it u-v} sampling are $\theta_{\mathrm{FWHM}} = \timeform{0.091"} \times \timeform{0.082"}, PA=\timeform{43D}$ in 2017 and $\theta_{\mathrm{FWHM}} = \timeform{0.150"} \times \timeform{0.091"}, PA=\timeform{87D}$ in 2021, respectively. The maximum detectable angular scale of this observation is about 10 times larger than the angular resolution.
The channel maps of  the SiO maser lines were made in the velocity range from $V_{\mathrm{LSR}}=-350$ km s$^{-1}$ to $V_{\mathrm{LSR}}=+350$ km s$^{-1}$ using the CASA task, {\tt tclean}.
To improve the sensitivity, the velocity integration width of the channel map is set to $\Delta V=5$ km s$^{-1}$. The integration width is too wide to study the maser emission from the stars itself. However, it is suitable to derive the radial motions of SiO maser stars.
The typical Root Mean Square (RMS) noise levels in the emission-free areas of the resulting maps are $1.1$ mJy/beam/$5$ km s$^{-1}$ in 2017 and $1.4$ mJy/beam/$5$ km s$^{-1}$ in 2021, respectively. 
These noise levels are the values at the center of the FOV.
The outer limit of the FOV was set so that the attenuation by the primary beam pattern of the 12 m antenna is 0.2 because the ALMA has high sensitivity. Therefore the radius of the FOV was widen up to $\sim\timeform{50"}$ or 2 pc. The noise levels at the outer limit are $5.5$ mJy/beam/$5$ km s$^{-1}$ in 2017 and $7.0$ mJy/beam/$5$ km s$^{-1}$ in 2021, respectively. 
%%%%%%%%%%%
 \begin{figure}
 \begin{center}
   \includegraphics[width=120mm]{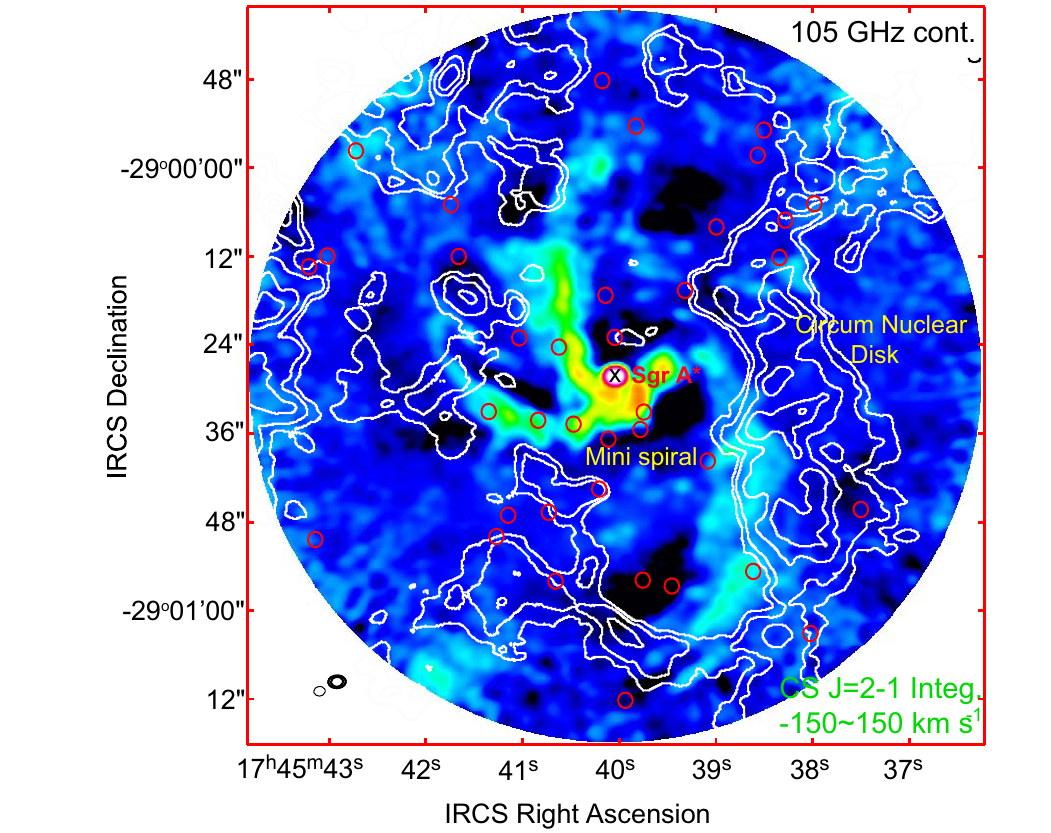}
 \end{center}
 \caption{Positions of detected stars in the SiO $v=1, J=2-1$ emission line  plotted on the continuum map at 105 GHz (pseudo color; \cite{Tsuboi2016}) and the integrated intensity map of the CS $J=2-1$ emission line  (contours; \cite{Tsuboi2018}). The red circles show the position of detected stars. The cross shows the position of Sgr A$^\ast$. The famous ionized gas flow around Sgr A$^\ast$, called the "Mini-Spiral (MS)", is the three-armed spiral-like feature on the continuum map. There is the "Circum Nuclear Disk (CND)" surrounding the MS, which is a molecular gas ring rotating around Sgr A$^\ast$.  There does not seem to be a large number of SiO maser stars in correlation with MS, CND,  and Sgr A$^\ast$. {Alt text: Circles shown positions of SiO maser stars on the two overlaid maps of CS line emission and 105 GHz continuum emission. }}
  \label{1}
\end{figure}
%%%%%%%%%%%%

 \section{Results}
 \subsection{Positions,  Line intensities, and LSR velocities of the SiO maser stars}
The weakness of the SiO $v=2, J=2-1$ emission line compared to the SiO $v=1, J=2-1$ emission line has been reported for late-type stars detected in the Galactic disk region (e.g. \cite{Bujarrabal}, \cite{Nakashima}, \cite{Lewis2024}).  Only the SiO $v=2, J=2-1$ emission line of IRS 7 (17454004-2900228) was detected in this sample.  It is confirmed that the SiO $v=2, J=2-1$ emission line is also much weaker than the SiO $v=1, J=2-1$ emission line in the Galactic center region. Therefore, we used only the latter for the astrometric analysis which will be mentioned in the following subsection. 
We detected 37 SiO maser stars in the channel maps of  the SiO $v=1, J=2-1$ emission line  in both 2017 and 2021. This detection number is significantly improved from the previous analysis using the same data, which detected 28 stars (\cite{Darling2023}, \cite{Paine2024}).  Then, the statistical analysis of the kinematics of SiO maser stars shown in the following section is possible due to the increase in detection. 
The positions in the ICRS coordinate system of the SiO maser  stars are determined by the 2D Gaussian fitting to the star image in the channel maps using the CASA task, {\tt imfit}. 
However, the positions of two SiO maser stars, 17453828-2900069 and 17453854-2859582, could not be determined in 2021 although they were determined  in 2017. These failures are caused by the line intensity decrease of the SiO maser emission in 2021. 

Figure 1 shows the positions in 2017 of the detected stars in the SiO $v=1, J=2-1$ emission line (red circles)  plotted on the continuum map at 105 GHz (pseudo color; \cite{Tsuboi2016}) and the integrated intensity map of the CS $J=2-1$ emission line (contours; \cite{Tsuboi2018}). The integration velocity range of the CS $J=2-1$ emission line is from $V_{\mathrm{LSR}}=-150$ km s$^{-1}$ to $V_{\mathrm{LSR}}=+150$ km s$^{-1}$.
The three-armed spiral-like feature on the continuum map is a famous ionized gas flow around Sgr A$^\ast$. It is called the "Mini-Spiral (MS)"(\cite{Eckers1983}, \cite{Lo}). There is also the "Circum Nuclear Disk (CND)" surrounding the MS in this figure, which is a molecular gas ring rotating around the GCBH (e.g. \cite{Tsuboi2018} ). 
The SiO maser stars are not only distributed on the MS, but also on the outside of the CND.
However, there does not seem to be a large number of SiO maser stars correlated with the MS, the CND, and Sgr A$^\ast$. 
The positions of the SiO maser stars in both 2017 and 2021 are summarized in Table 1.

%%%%%%%%%%%%%%%%%%%%%
\begin{figure}
 \begin{center}
  \includegraphics[width=90mm]{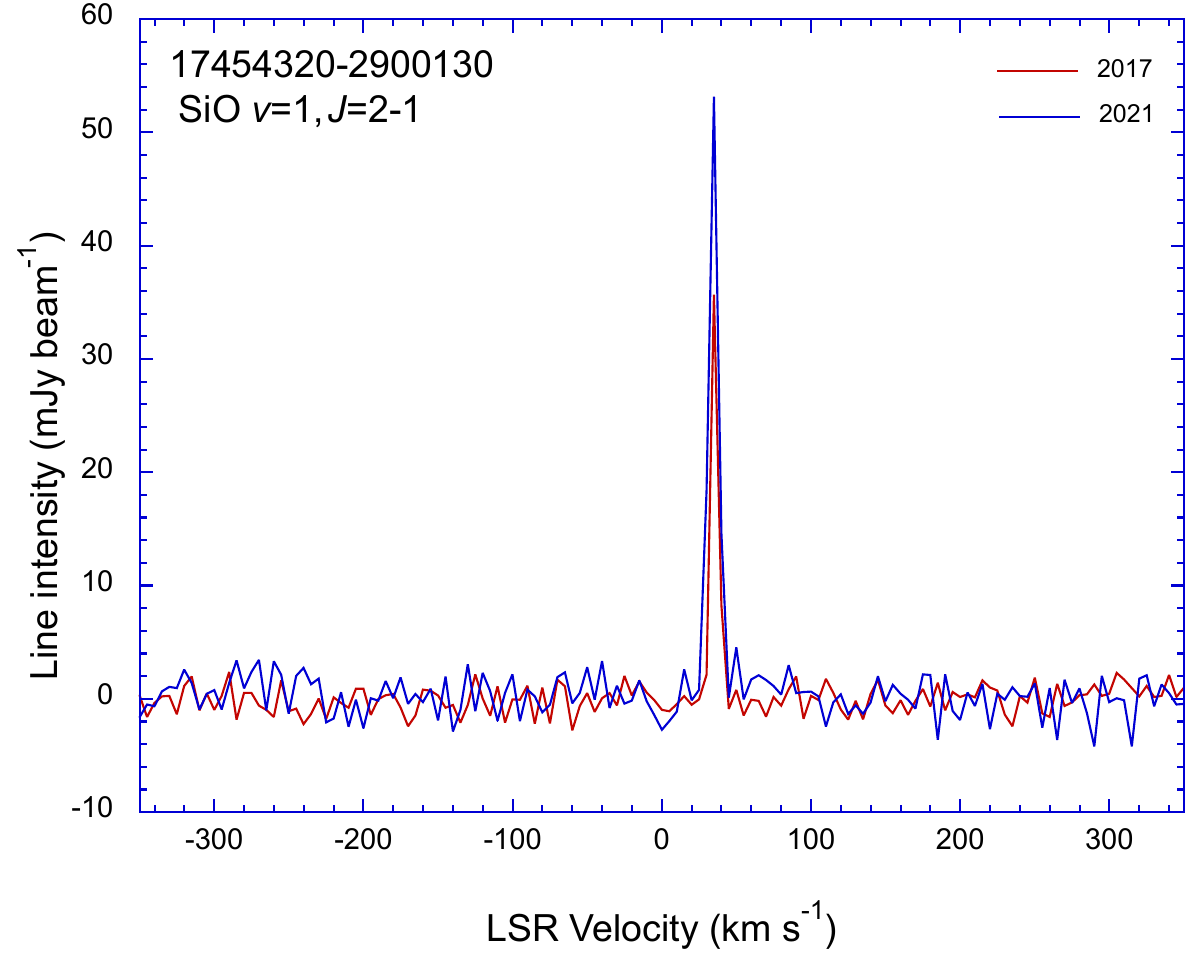}
 \end{center}
 \caption{Line profiles of the SiO $v=1, J=2-1$ emission line toward 17454320-2900130 as an example. The velocity interval  is 5 km s$^{-1}$.  All other line profiles are shown in Figure 10 in Appendix. {Alt text: Two line graphs showing the line profiles at 2017 and 2021  toward 17454320-2900130.}}
  \label{2}
\end{figure}
%%%%%%%%%%%%%%%%%%%%%

Figure 2 shows the line profiles of the SiO $v=1, J=2-1$ emission line towards a SiO maser star, 17454320-2900130, as an example.  We made the line profiles of  the SiO $v=1, J=2-1$ emission line from the channel maps and derived the LSR radial velocities from the line profiles with Gaussian fitting. To obtain the line intensity of the SiO $v=1, J=2-1$ emission line, the attenuation by the primary beam pattern at the angular distance from Sgr A$^\ast$, $\Delta\theta$, was corrected by the formula; 
\begin{equation}
\label{ }
S_{corr} = S_{obs} \exp[4\ln2(\Delta\theta/FWHM)^2]. 
\end{equation}
All other line profiles are shown in Figure 10 in Appendix.
Among the two SiO maser stars failed to obtain the 2021 positions, the line profiles in both 2017 and 2021 are easily identifiable for 17453828-2900069 while the 2021 line profile is barely identifiable for 17453854-2859582.

Figure 3 shows the line intensity evolution of the SiO $v=1, J=2-1$ emission line between 2017 and 2021. Although significant changes  are observed for all stars between two epochs, all stars are detected in both epochs as mentioned above. The line intensity of 19 stars decrease and those of 18 stars increase between two epochs. These observed changes are consistent with those of late type stars detected in the Galactic disk region, which is usually variable in several hundred days (e.g. \cite{Nakashima}, \cite{Lewis2024}).
On the other hand, the change of the LSR velocity is negligible for almost all stars.  
The LSR velocities of 15 stars have plus sign and those of  22 stars have minus sign. The mean LSR velocity is $\overline{V_\mathrm{LSR}} =-24.9\pm86.7$ km s$^{-1}$. There is no significant trend in the LSR velocities. 
%%%%
Note that the observed LSR velocities are those of the SiO maser spots around the stars, not the LSR velocities of the stars themselves. The difference can be as much as $\sim10$ km s$^{-1}$. However, this is not a major problem in this analysis.
%%%%
The line intensities and LSR velocities of the SiO maser stars are also summarized in Table 1.

%%%%%%%%%%%%%%%%%%%%%
\begin{figure}
 \begin{center}
  \includegraphics[width=80mm]{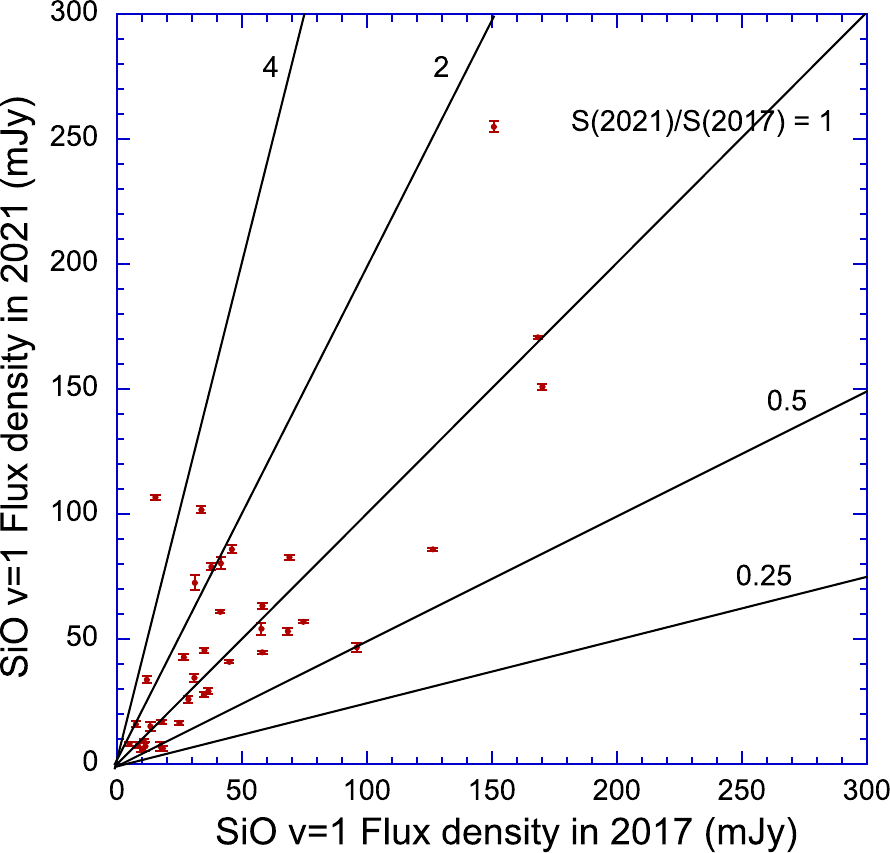}
  \end{center}
 \caption{Line intensity evolution of the SiO $v=1, J=2-1$ emission line between 2017 and 2021. The horizontal axis and the vertical axis are the line intensities in 2017 and 2019, respectively.  Linear lines show the ratios of the line intensities at two epochs.  {Alt text: A dispersion diagram. }
 }\label{3}
\end{figure}
%%%%%%%%%%%%%%%%%%%%%
\begin{figure}
 \begin{center}
  \includegraphics[width=160mm]{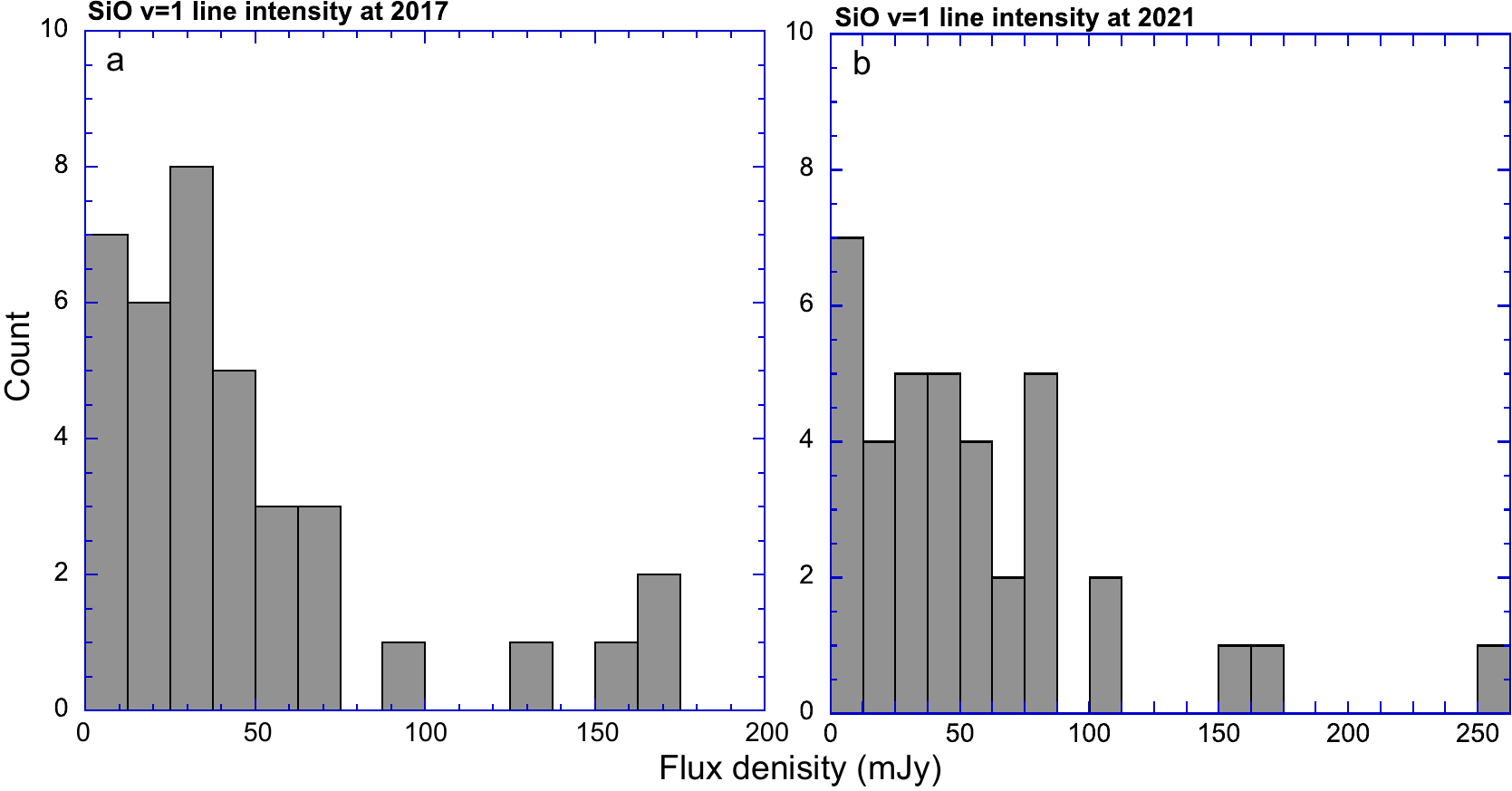}
 \end{center}
 \caption{{\bf a} Histogram of the line intensity of the SiO $v=1, J=2-1$ emission line of detected stars in 2017.
 The ratio of the detected stars with smaller than 50 mJy in 2017 is $\sim70\%$.  {\bf b} Histogram of the line intensity of the SiO $v=1, J=2-1$ emission line of detected stars in 2021. The ratio of the detected stars with smaller than 50 mJy in 2021 is $\sim60\%$. {Alt text: Two histograms. The horizontal axis and the vertical axis are the line intensity and count, respectively. }
 }\label{4}
\end{figure}
%%%%%%%%%%%%%%%%%%%%%

Figures 4a and 4b show the histograms of the line intensities in the SiO $v=1, J=2-1$ emission line of the detected stars in 2017 and 2021, respectively. 
The line intensities of the $\sim70\%$ of the detected stars are less than 50 mJy in 2017. Only four stars have line intensities greater than 100 mJy.   The average line intensities are $47.0\pm43.6$ mJy in 2017.  On the other hand, the line intensities of $\sim60\%$ of the detected stars are smaller than 50 mJy in 2021. Five stars have line intensities greater than 100 mJy. The average line intensities are $55.4\pm52.0$ mJy in 2021. The situation has not changed between two epochs.
Most of them could be observed as inconspicuous maser stars with line intensities of a few Jy when they were within a few kpc (Cf. \cite{Lewis2024}).

Figure 5 shows the radial distribution of the SiO $v=1, J=2-1$ emission line intensity of the detected stars in 2017 as a function of the angular distance from Sgr A$^\ast$, whose derivation will be mentioned in the next subsection.
There is no clear trend in the radial distribution. The stars with $S >100$ mJy are widely scattered. The stars with $10 < S < 50$ mJy were detected around Sgr A$^\ast$ and also near the outer edge of the channel map. On the other hand, the star with $S < 10$ mJy was not detected near the outer edge of the channel map. 
Attenuation by the primary beam pattern causes the detection efficiency to decrease somewhat near the outer edge. 
However, this could not affect the astrometric analysis mentioned in the following subsection. 

%%%%%%%%%%%%%%%%%%%%%
\begin{figure}
 \begin{center}
  \includegraphics[width=80mm]{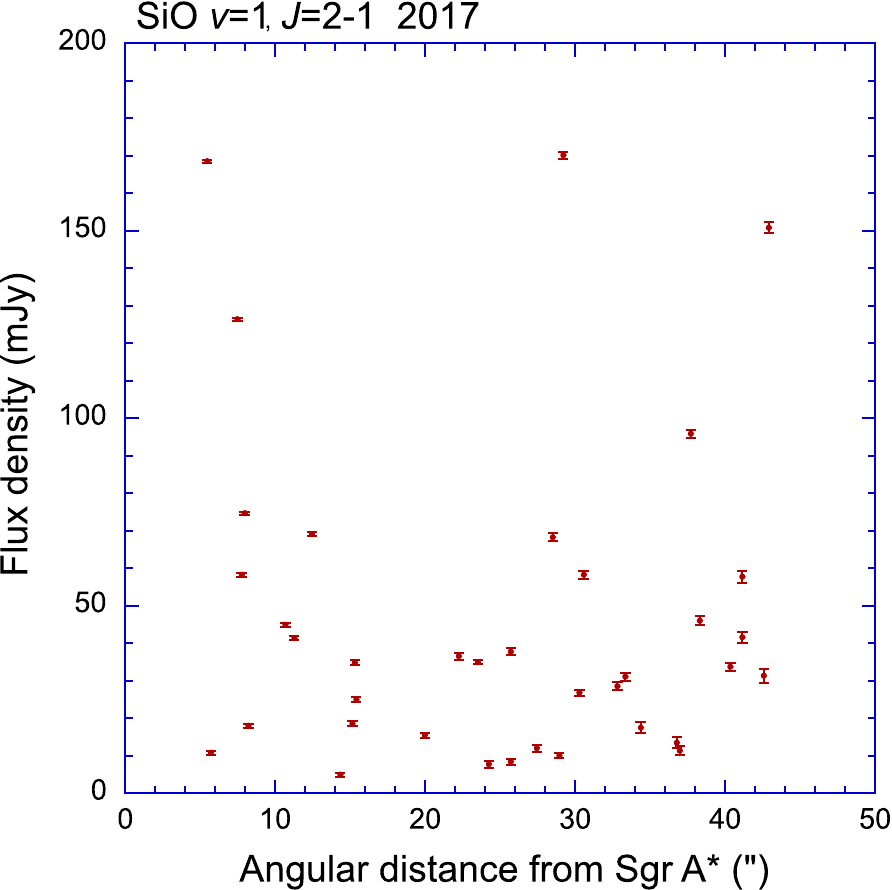}
  \end{center}
 \caption{Radial distribution of the SiO $v=1, J=2-1$ emission line intensity of detected stars in 2017 on the angular distance from Sgr A$^\ast$. {Alt text: A dispersion diagram. The horizontal axis and the vertical axis are the angular distance from Sgr A$^\ast$ and line intensity, respectively.}
 }\label{5}
\end{figure}
%%%%%%%%%%%%%%%%%%%%%
%%%%%%%%%%%%%%%%%%%%%
\begin{figure}
  \begin{center}
\includegraphics[width=170mm]{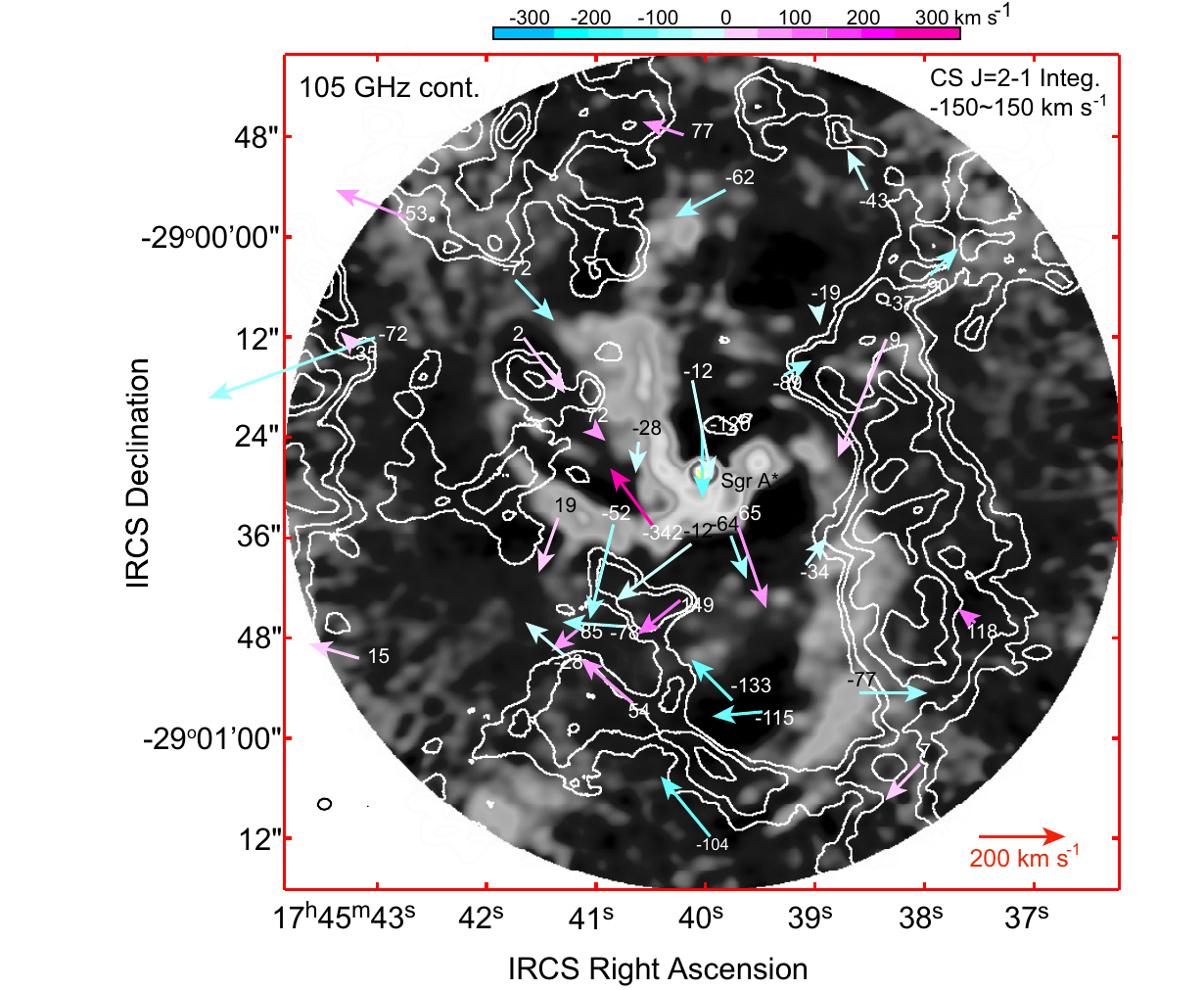}
 \end{center}
 \caption{Proper motions in vectors and LSR radial velocities in color of the SiO maser stars around Sgr A$^\ast$ plotted on  the continuum map at 105 GHz (black and white; \cite{Tsuboi2016}) and the integrated intensity map of the CS $J=2-1$ emission line  (contours; \cite{Tsuboi2018}).
 The lengths of the vectors indicate the amplitudes of the proper motions. The numbers to the sides of the vectors indicate the LSR velocities. {Alt text: Vectors shown proper motions of SiO maser stars on the two overlaid maps of CS line emission and 105 GHz continuum emission. }}
 \label{6}
\end{figure}
%%%%%%%%%%%%%%%%%%%%%

\subsection{Proper motions  of the SiO maser stars}
Since the position of Sgr A$^\ast$ in the ICRS coordinate system changes with the Galactic rotation of the Sun (e.g  \cite{Reid}, \cite{Xu2022}, \cite{Oyama}), the angular differences between the positions of the SiO maser stars summarized in Table 1 are not their proper motions. 
On the continuum maps mentioned above, the positions of Sgr A$^\ast$ in the ICRS coordinate system were measured to be (RA, Dec)$_{\rm ICRS}$ = (\timeform{17h45m40.033456s}\timeform{\pm0.000002s}, \timeform{-29D00'28.22993''}\timeform{\pm 0.00002"}) at September 19, 2017  
and  (RA, Dec)$_{\rm ICRS}$ = (\timeform{17h45m40.032847s}\timeform{\pm0.000005s}, \timeform{-29D00'28.26277"}\timeform{\pm0.00003''}) at August 19, 2021, respectively.
We obtained the positions relative to Sgr A$^\ast$, ($\Delta$RA, $\Delta$Dec), by subtracting these positions of Sgr A$^\ast$ from the positions shown in Table 1 using 2D Gaussian fitting, the CASA task, {\tt imfit}. The positions relative to Sgr A$^\ast$ of the SiO maser stars are summarized in Table 2.

We obtained  the angular differences between their positions relative to Sgr A$^\ast$ in 2017 and 2021 in Cartesian coordinate system,  which are given by 
\begin{equation}
\label{ }
\Delta X = -(\Delta\mathrm{RA}_{2021}-\Delta\mathrm{RA}_{2017})
\end{equation}
%and 
 \begin{equation}
\label{ }
\Delta Y= \Delta\mathrm{Dec}_{2021}-\Delta\mathrm{Dec}_{2017}.
\end{equation}
Then we determined the proper motions relative to Sgr A$^\ast$ of 35 SiO maser stars using the following formula;
\begin{equation}
\label{ }
|V_{\mathrm{prop}}| [\mathrm{km~ s}^{-1}] = \frac{38827}{\Delta t}\sqrt{\Delta X^2+\Delta Y^2}.
\end{equation}
The time difference between the observation epochs is $\Delta t = 3.917$ years. 
Because the origin and rotation direction of the position angle are defined as north and counterclockwise, respectively, the different formula is used for the derivation in west ($\Delta X > 0$) and east ($\Delta X < 0$) .
These are 
\begin{equation}
\label{ }
PA [\timeform{D}] = -90+\arctan{\frac{\Delta Y}{\Delta X}}~~(\Delta X > 0) 
\end{equation}
and
\begin{equation}
\label{ }
PA [\timeform{D}] = 90+\arctan{\frac{\Delta Y}{\Delta X}}~~(\Delta X < 0).
\end{equation}
The derived proper motions of stars detected in previous observations, such as that of IRS7 (17454004-2900228), are consistent with the previously derived proper motions (e.g. \cite{Tsuboi2022}, \cite{Darling2023}, \cite{Paine2024}). 
The proper motions relative to Sgr A$^\ast$ of the SiO maser stars are also summarized in Table 2.

Figure 6 shows the proper motions relative to Sgr A$^\ast$ in vectors and the LSR radial velocities in color of the SiO maser stars plotted on  the continuum map at 105 GHz (black and white; \cite{Tsuboi2016}) and the integrated intensity map of the CS $J=2-1$ emission line  (contours; \cite{Tsuboi2018}).
The derived proper motions of the SiO maser stars do not look completely random. 
If anything, some co-moving groups can be identified (Cf. Fig. 7 in \cite{Tsuboi2022}). 
The ionized gas streams of the MS have been reported to orbit the GCBH on high eccentricity Keplerian orbits \citep{Zhao2009}. 
The derived proper motions of the SiO maser stars do not appear to be such Keplerian orbits. 
While the CND has been reported to orbit the GCBH on a nearly circular orbit (e.g. \cite{Tsuboi2018}), the proper motions do not appear to be such a circular orbit. 
Therefore, the proper motions are not correlated with the MS and the CND.

%\onecolumn
\begin{figure}
 \begin{center}
  \includegraphics[width=160mm]{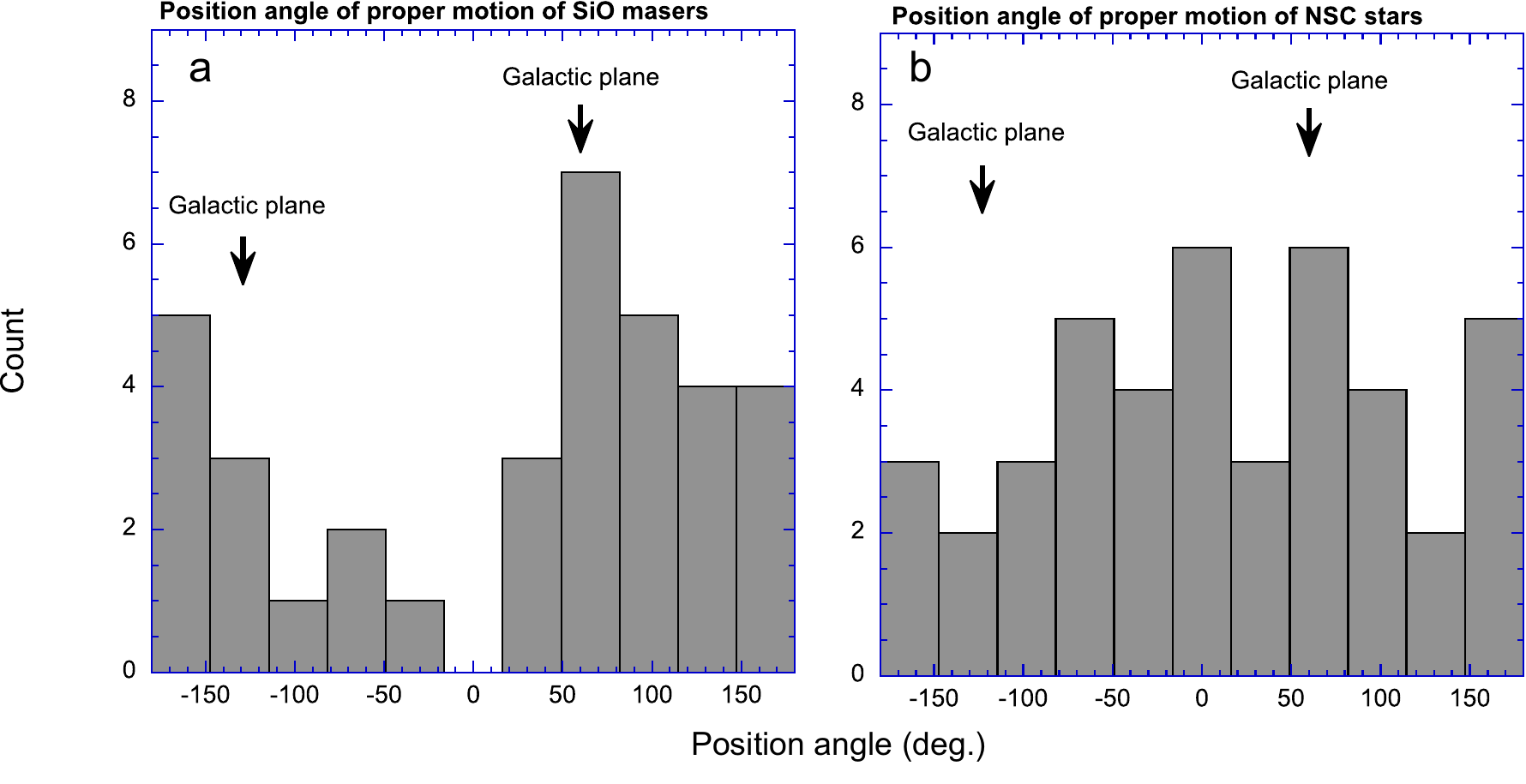}
 \end{center}
  \caption{\textbf{a} Histogram of the position angle of the proper motions relative to Sgr A$^\ast$ of 35  SiO maser stars.
  The black arrows indicate the direction of Galactic plane. \textbf{b} Histogram of the position angle of the proper motions of the WR and O stars in the NSC, excluding the co-moving clusters IRS13E and IRS13N (Tsuboi et al. 2022). {Alt text: Two histograms. The horizontal axis and the vertical axis are the position angle and count, respectively.}}
\label{7}
\end{figure}
%\twocolumn
%\onecolumn
\begin{figure}
 \begin{center}
  \includegraphics[width=160mm]{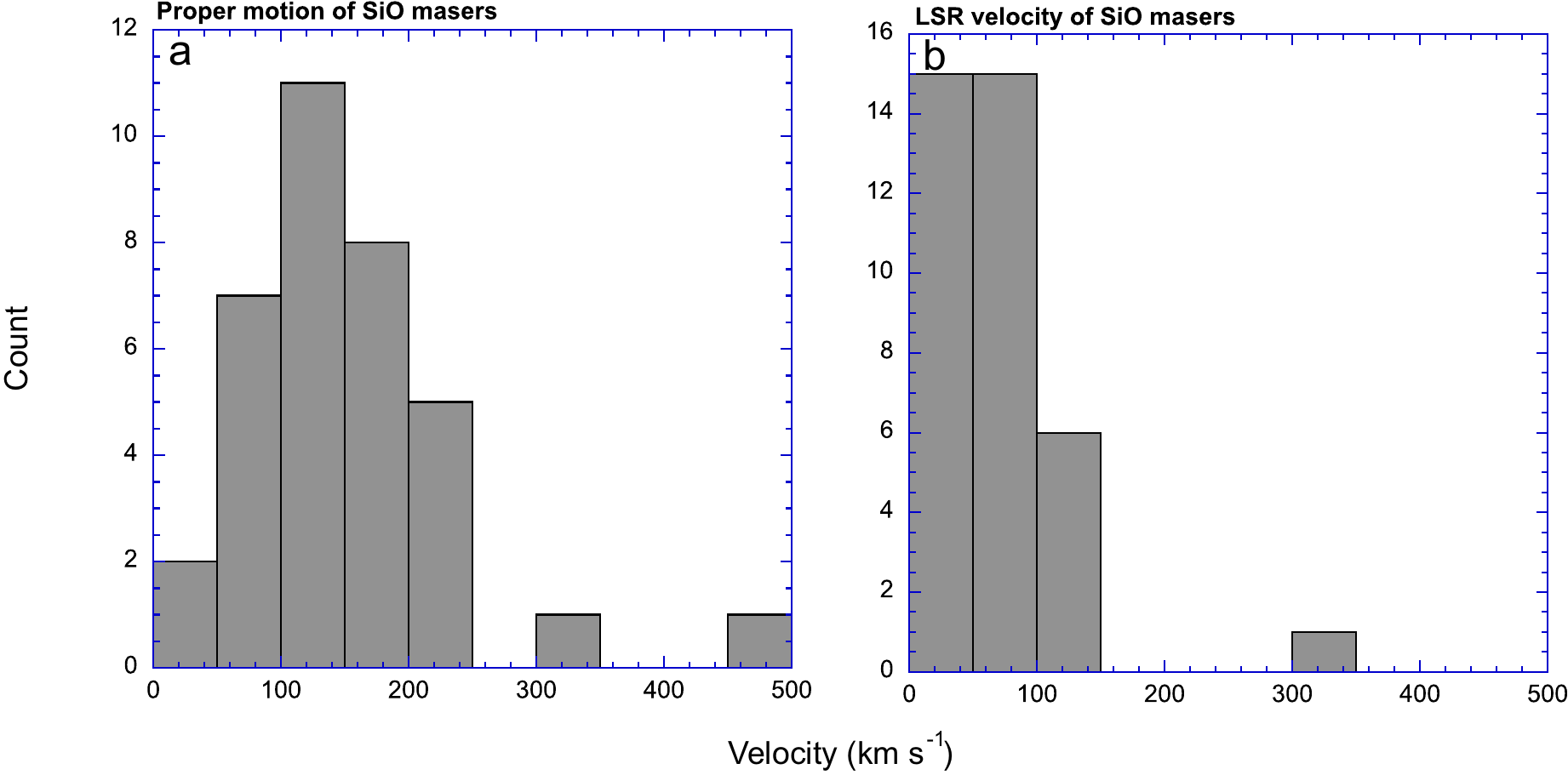}
 \end{center}
  \caption{ \textbf{a} Histogram of the proper motion amplitudeof SiO maser stars. The mean proper motion amplitude is $\overline{|V_\mathrm{prop}|} =153.2\pm83.3$ km s$^{-1}$.  \textbf{b} Histogram of the LSR velocity amplitude of SiO maser stars. The mean amplitude of the LSR velocity is $\overline{|V_\mathrm{LSR}|} =66.1\pm60.6$ km s$^{-1}$. {Alt text: Two histograms. The horizontal axis and the vertical axis are the velocity and count, respectively.}}
\label{8}
\end{figure}
%\twocolumn
%%%%%%%
%%%%%%%%%%%%%%%%%%%%%
\begin{figure}
 \begin{center}
  \includegraphics[width=80mm]{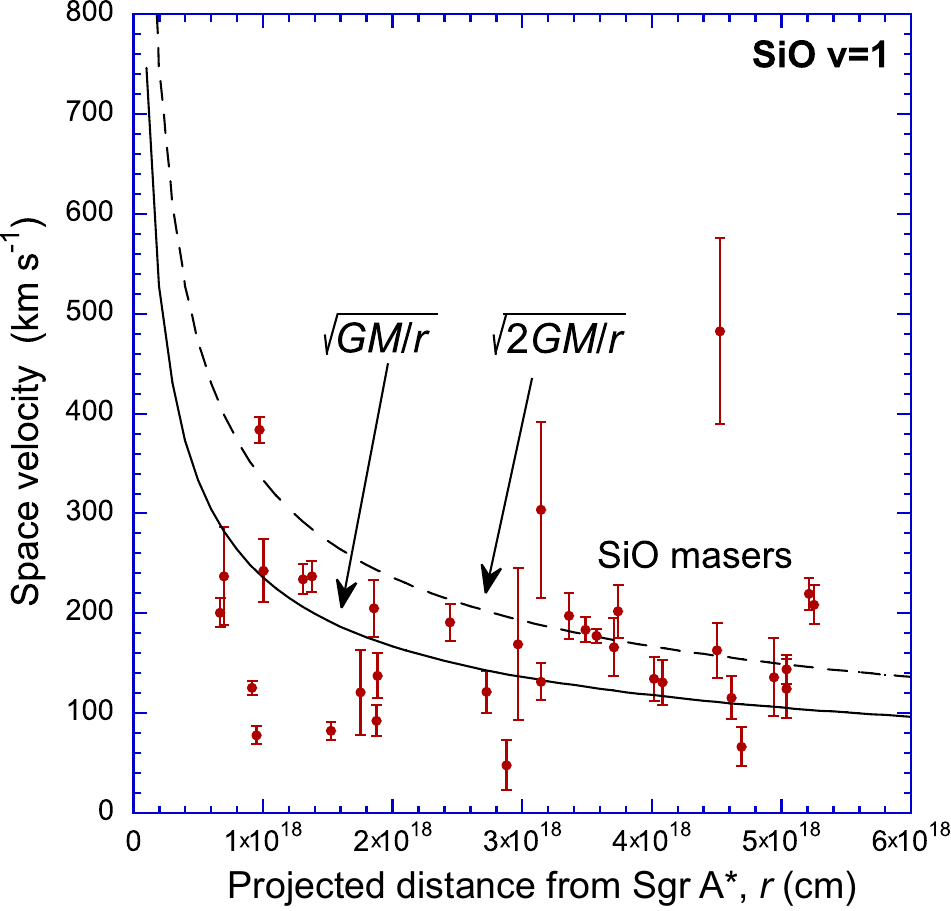}
 \end{center}
 \caption{Space velocities of the SiO maser stars at projected distances, $r$, from Sgr A$^\ast$.   The solid line curve indicates the orbital velocity of circular orbits, $V=\sqrt{\frac{GM}{r}}$, assuming that the mass of the GCBH is $M = 4\times10^6 M_\odot$. The dashed curve indicates the orbital velocity of extremely highly eccentric orbits, $V=\sqrt{\frac{2GM}{r}}$. {Alt text: A dispersion diagram.} }
 \label{9}
\end{figure}
%%%%%%%%%%%%%%%%%%%%%

 \section{Discussion}
\subsection{Proper motion along the Galactic plane}
As mentioned above, we determined the proper motions relative to Sgr A$^\ast$ of 35 SiO maser stars.
Figure 7a shows the histogram of the position angles of the proper motion relative to Sgr A$^\ast$ of the SiO maser stars. 
The distribution of the position angles are bipolar rather than random.
The black arrows indicate the direction of the Galactic plane. The peaks of the distribution coincide with the direction of the Galactic plane. That is, the proper motions of the SiO maser stars show a tendency to lie along the Galactic plane.
For comparison, Figure 7b shows the histogram of the position angle of the proper motions of the WR and O stars in the NSC, excluding the co-moving clusters IRS13E and IRS13N \citep{Tsuboi2022}. The black arrows also indicate the direction of the galactic plane.
The proper motions of the NSC stars are rather random. A tendency along the Galactic plane is not apparent. 
It is therefore plausible that the statistical property of the proper motions of the SiO maser stars is different from that of the NSC.

If the SiO maser stars belonged to the NSC, their motions were well randomized by a short time scale in the NSC, for example the period of the Keplerian orbit with the radius of 2 pc is $P\sim 1.3\times10^5$  years. As the result, they should have a similar statistical property  of the proper motions. On the other hand, if the SiO maser stars belonged to the NSD, the position angles were such that they were observed. This situation is consistent with that the proper motions of the SiO maser stars are not correlated with the MS and the CND (see Results).
 
 \subsection{Disproportionate proper motion speed and LSR speed}
Figures 8a and 8b show the histograms of the proper motion amplitude and the LSR velocity amplitude of SiO maser stars, respectively. 
The mean proper motion amplitude is derived to be $\overline{|V_\mathrm{prop}|} =153.2\pm83.3$ km s$^{-1}$ (Cf. $\overline{|V_\mathrm{prop}|} =239.4\pm127.7$ km s$^{-1}$ for the NSC stars in \cite{Tsuboi2022}). While the mean amplitude of the LSR velocity is derived to be $\overline{|V_\mathrm{LSR}|} =66.1\pm60.6$ km s$^{-1}$.
 The proper motion amplitude is larger than the LSR velocity amplitude. 
That is, in the observed SiO maser stars, the motions across the line of sight dominate the motions along the line of sight.  Moreover the motions have  a tendency to be along the Galactic plane (see Figure 7a). 

If the SiO maser stars belonged to the NSC,  their motions would be well randomized (see the previous subsection) and  both amplitudes should be similar. 
On the other hand, if the SiO maser stars belong to the NSD (see \cite{Sanders2022}), they would pass in front of or behind the GCBH on orbits with larger radii.
Then their motions should be such that they could be observed here. 
%%%%%%%%%%%%%%%%%%%%%%%%%%%%%%%%%%%%%%% %%%%%%%%%%%%%%%%%%%%%%%%%%%%%%%%%%%%%
 \subsection{Large enclosed mass}
We derived the space velocities  of the SiO maser stars from these proper motions and the LSR radial velocities using $V_{\mathrm{space}} = \sqrt{V_{\mathrm{prop.}}^2+V_{\mathrm{LSR}}^2} $.
Figure 9 shows the space velocities of the SiO maser stars at projected distances, $r$, from Sgr A$^\ast$. 
There is no clear tendency for stars near Sgr A$^\ast$ to have large velocities and stars far from Sgr A$^\ast$ to have small velocities, as observed in the NSC. The solid line curve indicates the orbital velocity of circular orbits, $V=\sqrt{\frac{GM}{r}}$, assuming that the mass of the GCBH is $M = 4\times10^6 M_\odot$. 
The dashed curve indicates the orbital velocity of extremely highly eccentric orbits, $V=\sqrt{\frac{2GM}{r}}$.  Many inferred space velocities are larger than the former. Several reach or exceed the latter (Cf. Fig.8 in \cite{Tsuboi2022}). 

If the stars belong to the NSC, the $V=\sqrt{\frac{2GM}{r}}$ should be the upper limit of the space velocities.
To resolve the discrepancy with the observed space velocities, it is hypothesized that there may be additional mass in the region.  Since the mass appears to accelerate the velocities of the stars by up to several 10 \%, the required mass is expected to be as large as $M \sim 10^6 M_\odot$. However, such a mass has not been observed in other observations (e.g.  \cite{Reid}, \cite{Oyama}).
If the SiO maser stars belong to the NSD, this kinematics can be expected.

\section{Conclusions}
We present the kinematics of SiO maser stars around Sgr A$^\ast$ using the archive data in the SiO $v=1, J=2-1$ emission line obtained by the Atacama Large Millimeter/Submillimeter Array (ALMA) in 2017 and 2021.  
\begin{itemize}
  \item We detected 37 SiO maser stars and derived their angular offsets relative to Sgr A$^\ast$ and LSR radial velocities. 
  \item We derived the proper motions of 35 SiO maser stars by comparing their angular offsets in the two epochs. 
  \item The proper motions of the SiO maser stars show a tendency to lie along the Galactic plane.
  \item The proper motion amplitudes are  larger than the LSR velocity amplitude. 
  \item Many 3D velocities are near to or larger than the upper limit velocities for Kepler orbits around Sgr A$^\ast$, whose mass is assumed to be $4\times10^6 M_\odot$.
  \item  These results indicate that the SiO maser stars are members of the NSD rather than the NSC. 
\end{itemize}

%%%%%%%%%%%%%%%%%%%%%%%%%%%%%%%%%%%%%%

\begin{ack}
 ALMA is a partnership of European Southern Observatory (ESO) (representing its member states), the National Science Foundation (NSF) (USA) and National Institutes of Natural Sciences (NINS) (Japan), together with National Research Council Canada (NRC) (Canada), Ministry of Science and Technology (MOST) and Institute of Astronomy and Astrophysics, Academia Sinica (ASIAA) (Taiwan), and Korea Astronomy and Space Science Institute (KASI) (Republic of Korea), in cooperation with the Republic of Chile. 
The National Radio Astronomy Observatory (NRAO) is a facility of NSF operated under cooperative agreement by Associated Universities, Inc. USA (AUI). 
 The Joint ALMA Observatory (JAO) is operated by ESO, AUI/NRAO and National Astronomical Observatory of Japan (NAOJ). 
  This paper makes use of the following ALMA data ADS/JAO. ALMA\#2016.1. 00940.S and ALMA\#2019.1.00292. 
\end{ack}

%%%%%%%%%%%%%%%%%%%%%%%%%%%%%%%%%%%%%%

%\clearpage
\small\begin{table*}[p]
\centering
\rotatebox{90}{\begin{minipage}{\textwidth} 
%\begin{table}
  \tbl{Positions,  line intensities, and LSR velocities of the SiO maser stars. }{%
  \begin{tabular}{ccccccccc}
  \hline
No.&Name &RA in 2017\footnotemark[$\dag$]&Dec in 2017\footnotemark[$\dag$]&RA in 2021\footnotemark[$\dag$]&Dec in 2021\footnotemark[$\dag$]&Line intensity\footnotemark[$\ast$]& Line intensity\footnotemark[$\ast$]& LSR velocity\footnotemark[$\ddag$]  \\ 
&&\timeform{17h45m}+ ~ [\timeform{s}]&\timeform{-29D00'}+ ~ [\timeform{''}]&\timeform{17h45m}+  ~[\timeform{s}]&\timeform{-29D00'}+ ~ [\timeform{''}]& in 2017~[mJy] & in 2021~[mJy]&[km s$^{-1}$] \\ 
\hline 
1&17453751-2900462&$37.50992\pm0.00009$&$-46.15660\pm0.00126$&$37.50975\pm0.00011$&$-46.18955\pm0.00057$& $31.1\pm1.0$&$ 34.6\pm1.6 $& $117.7\pm0.4$\\ 
2&17453799-2900046&$37.98760\pm0.00007$&$-4.62436\pm0.00094$&$37.98639\pm0.00012$&$-4.65102\pm0.00074$& $28.5\pm1.1$&$ 25.9\pm1.6 $& $-90.3\pm0.7$\\ 
3&17453802-2901027&$38.01949\pm0.00011$&$-62.70482\pm0.00167$&$38.01959\pm0.00014$&$-62.74611\pm0.00069$& $57.7\pm1.6$&$ 54.2\pm2.4$& $7.2\pm0.2$\\ 
4&17453828-2900069&$38.27841\pm0.00049$&$-6.91983\pm0.00544$&-&-& $10.0\pm0.7$&$ 5.9\pm1.1$& $-37.2\pm0.7$\\
5&17453830-2900113&$38.29962\pm0.00024$&$-11.2917\pm0.00353$&$38.30002\pm0.00052$&$-11.35216\pm0.00328$& $8.4\pm0.8$&$ 7.6\pm1.1$& $9.1\pm5.8$\\
6&17453852-2859545&$38.51369\pm0.00006$&$5.44986\pm0.00072$&$38.51349\pm0.00013$&$5.42638\pm0.00087$& $95.9\pm1.1$&$ 46.8\pm1.8$& $-43.1\pm0.2$\\
7&17453854-2859582&$38.53883\pm0.00019$&$1.81872\pm0.00155$&-&-& $17.4\pm1.5$&$ 7.1\pm1.8$& $10.0\pm0.8$\footnotemark[$\bot$]\\ 
8& 17453861-2900543&$38.60795\pm0.00006$&$-54.32691\pm0.00085$&$38.60590\pm0.00017$&$-54.35982\pm0.00080$& $58.3\pm1.0$&$ 63.3\pm1.3$& $-76.7\pm0.3$\\
9&17453898-2900079&$38.97678\pm0.00008$&$-7.87575\pm0.00108$&$38.97611\pm0.00014$&$-7.91295\pm0.00068$& $35.0\pm0.6$&$45.6\pm1.1$& $-18.8\pm4.1$\\
10&17453908-2900393&$39.08295\pm0.00004$&$-39.29686\pm0.00052$&$39.08190\pm0.00010$&$-39.32330\pm0.00052$&$34.8\pm0.7$&$27.9\pm1.1$& $-34.4\pm0.4$\\
11&17453932-2900164&$39.31030\pm0.00018$&$-16.38664\pm0.00170$&$39.30912\pm0.00022$&$-16.41605\pm0.00133$&$4.9\pm0.6$&$8.1\pm0.9$&$-89.0\pm1.1$\\
12&17453945-2900567&$39.45215\pm0.00003$&$-56.70001\pm0.00036$&$39.45258\pm0.00003$&$-56.73391\pm0.00016$&$170.1\pm1.0$&$150.9\pm1.4$&$-114.8\pm0.1$\\
13& 17453975-2900330&$39.75042\pm0.00013$&$-32.98933\pm0.00146$&$39.74915\pm0.00032$&$-33.04350\pm0.00159$&$10.8\pm0.6$&$9.2\pm0.9$&$65.0\pm1.1$\\
14& 17453977-2900555&$39.75396\pm0.00010$&$-55.49524\pm0.00082$&$39.75421\pm0.00012$&$-55.51854\pm0.00051$&$12.1\pm0.9$&$33.9\pm1.4$&$-132.8\pm0.7$\\
15& 17453979-2900352&$39.78378\pm0.00002$&$-35.17128\pm0.00028$&$39.78284\pm0.00004$&$-35.21413\pm0.00019$&$126.3\pm0.5$&$86.0\pm0.7$&$-63.7\pm0.2$\\
16& 17453995-2901108&$39.94840\pm0.00006$&$-70.79247\pm0.00095$&$39.94883\pm0.00007$&$-70.81127\pm0.00030$&$31.3\pm1.8$&$72.6\pm3.0$&$-103.6\pm0.6$\\
17& 17453996-2859515&$39.95894\pm0.00008$&$8.54725\pm0.00070$&$39.95939\pm0.00016$&$8.50823\pm0.00133$&$13.5\pm1.4$&$15.2\pm1.8$&$-62.4\pm1.3$\\
18& 17454004-2900228&$40.03598\pm0.00003$&$-22.76024\pm0.00038$&$40.03535\pm0.00010$&$-22.80930\pm0.00045$&$168.5\pm0.4$&$170.8\pm0.6$&$-119.7\pm0.4$\\
19& 17454011-2900364&$40.10550\pm0.00013$&$-36.39556\pm0.00156$&$40.10646\pm0.00015$&$-36.44162\pm0.00082$&$18.0\pm0.5$&$17.0\pm0.8$&$-12.3\pm0.4$\\
20& 17454013-2900170&$40.12410\pm0.00005$&$-17.02117\pm0.00066$&$40.12307\pm0.00009$&$-17.07725\pm0.00044$&$41.3\pm0.5$&$61.1\pm0.7$&$-12.5\pm0.1$\\
21& 17454017-2859479&$40.16788\pm0.00020$&$12.08796\pm0.00248$&$40.16814\pm0.00011$&$12.05818\pm0.00054$&$33.7\pm1.1$&$101.9\pm1.5$&$70.7\pm0.4$\\
22& 17454021-2900432&$40.21000\pm0.00013$&$-43.24652\pm0.00154$&$40.21029\pm0.00010$&$-43.28735\pm0.00116$&$18.5\pm0.6$&$6.4\pm1.0$&$148.6\pm2.0$\\
23& 17454047-2900345&$40.46869\pm0.00003$&$-34.53782\pm0.00048$&$40.46897\pm0.00007$&$-34.55745\pm0.00036$&$74.7\pm0.4$&$57.0\pm0.6$&$-342.0\pm4.7$\\
24& 17454062-2900241&$40.61921\pm0.00004$&$-24.05349\pm0.00055$&$40.61866\pm0.00003$&$-24.09361\pm0.00014$&$58.2\pm0.5$&$44.8\pm0.6$&$-28.2\pm0.1$\\
25& 17454065-2900559&$40.65006\pm0.00003$&$-55.88373\pm0.00042$&$40.65054\pm0.00008$&$-55.90616\pm0.00043$&$68.3\pm1.0$&$53.1\pm1.5$&$53.8\pm0.5$\\
26& 17454075-2900465&$40.74547\pm0.00009$&$-46.55757\pm0.00098$&$40.74621\pm0.00006$&$-46.58941\pm0.00029$&$15.5\pm0.7$&$106.8\pm1.0$&$-77.5\pm6.3$\\
27& 17454083-2900341&$40.83352\pm0.00006$&$-34.11774\pm0.00073$&$40.83341\pm0.00008$&$-34.17266\pm0.00033$&$44.9\pm0.4$&$41.1\pm0.6$&$-52.2\pm0.8$\\
 \hline
\end{tabular}}\label{tab:first}
\begin{tabnote}
\footnotemark[$\dag$] Positions in the ICRS coordinate system. \\
\footnotemark[$*$] Integrated intensity of SiO $v=1, J=2-1$ emission line. \\
\footnotemark[$\ddag$] Mean of LSR velocities of SiO $v=1, J=2-1$ emission line in 2017 and 2021. \\
\footnotemark[$\bot$] Derived from the line profile only in 2017.
\end{tabnote}
%\end{table}
\end{minipage}}\end{table*}
\normalsize

%%%%%%%%%%%%%%%%%%%%%%%%%%%%%%%%%%%%%%% %%%%%%%%%%%%%%%%%%%%%%%%%%%%%%%%%%%%%
%%%%%%%%%%%%%%%%%%%%%%%%%%%%%%%%%%%%%%

\addtocounter{table}{-1}
%\clearpage
\small\begin{table*}[p]
\centering
\rotatebox{90}{\begin{minipage}{\textwidth} 
%\begin{table}
  \tbl{Continued. }{%
  \begin{tabular}{ccccccccc}
  \hline
No.&Name &RA in 2017\footnotemark[$\dag$]&Dec in 2017\footnotemark[$\dag$]&RA in 2021\footnotemark[$\dag$]&Dec in 2021\footnotemark[$\dag$]&Line intensity\footnotemark[$\ast$]& Line intensity\footnotemark[$\ast$]& LSR velocity\footnotemark[$\ddag$]  \\ 
&&\timeform{17h45m}+ ~ [\timeform{s}]&\timeform{-29D00'}+ ~ [\timeform{''}]&\timeform{17h45m}+  ~[\timeform{s}]&\timeform{-29D00'}+ ~ [\timeform{''}]& in 2017~[mJy] & in 2021~[mJy]&[km s$^{-1}$] \\ 
\hline 
28& 17454103-2900227&$41.03431\pm0.00003$&$-22.68209\pm0.00047$&$41.03347\pm0.00004$&$-22.71744\pm0.00021$&$69.1\pm0.5$&$82.7\pm0.9$&$72.2\pm0.2$\\
29& 17454115-2900467&$41.14440\pm0.00010$&$-46.70189\pm0.00124$&$41.14434\pm0.00008$&$-46.73965\pm0.00038$&$36.5\pm0.8$&$29.2\pm1.1$&$85.0\pm2.0$\\
30& 17454127-2900500&$41.27269\pm0.00008$&$-49.89565\pm0.00111$&$41.27288\pm0.00008$&$-49.92095\pm0.00035$&$37.8\pm0.9$&$79.0\pm1.4$&$-28.2\pm0.3$\\
31& 17454135-2900331&$41.34494\pm0.00009$&$-33.05335\pm0.00113$&$41.34474\pm0.00011$&$-33.09882\pm0.00055$&$24.9\pm0.6$&$16.6\pm0.9$&$18.5\pm0.4$\\
32& 17454172-2900129&$41.72159\pm0.00023$&$-12.92978\pm0.00455$&$41.72013\pm0.00035$&$-12.97552\pm0.00294$&$7.7\pm0.9$&$16.0\pm1.2$&$2.0\pm2.0$\\
33& 17454175-2900047&$41.74864\pm0.00008$&$-4.75022\pm0.00109$&$41.74718\pm0.00018$&$-4.793240\pm0.00084$&$26.7\pm0.9$&$42.7\pm1.2$&$-71.9\pm0.4$\\
34& 17454272-2859575&$42.72239\pm0.00008$&$2.47401\pm0.00104$&$42.72326\pm0.00009$&$2.447455\pm0.00040$&$150.9\pm1.6$&$254.9\pm2.2$&$52.6\pm0.4$\\
35& 17454303-2900120&$43.01342\pm0.00018$&$-11.98445\pm0.00173$&$43.01632\pm0.00068$&$-12.03141\pm0.00080$&$11.4\pm1.3$&$7.1\pm1.8$&$-72.4\pm2.4$\\
36& 17454315-2900503&$43.15197\pm0.00007$&$-50.27589\pm0.00111$&$43.15243\pm0.00003$&$-50.30538\pm0.00013$&$41.6\pm1.5$&$80.4\pm2.5$&$14.8\pm0.4$\\
37& 17454319-2900130&$43.18993\pm0.00009$&$-13.04773\pm0.00108$&$43.18966\pm0.00008$&$-13.07710\pm0.00041$&$46.0\pm1.2$&$86.1\pm1.6$&$35.3\pm0.3$\\
 \hline
\end{tabular}}\label{tab:first}
\begin{tabnote}
\footnotemark[$\dag$] Positions in the ICRS coordinate system. \\
\footnotemark[$*$] Integrated intensity of SiO $v=1, J=2-1$ emission line. \\
\footnotemark[$\ddag$] Mean of LSR velocities of SiO $v=1, J=2-1$ emission line in 2017 and 2021. \\
\footnotemark[$\bot$] Derived from the line profile only in 2017.
\end{tabnote}
%\end{table}
\end{minipage}}\end{table*}
\normalsize

%%%%%%%%%%%%%%%%%%%%%%%%%%%%%%%%%%%%%%% %%%%%%%%%%%%%%%%%%%%%%%%%%%%%%%%%%%%%
%\clearpage
\small\begin{table*}[p]
\centering
\rotatebox{90}{\begin{minipage}{\textwidth} 
%\begin{table}
  \tbl{Proper motions of the SiO maser stars. }{%
 \begin{tabular}{ccccccccc}
%\begin{tabular}{rrrrrrrrr}
  \hline
  No&$\Delta$RA relative to&$\Delta$RA realtive to&$\Delta X$\footnotemark[$\ddag$]&$\Delta$Dec relative to&$\Delta$Dec relative to&$\Delta Y$\footnotemark[$\ddag$]&$|V_{\mathrm{prop}}|$\footnotemark[$\bot$]&$PA$\\ 
&Sgr A$^\ast$ in 2017\footnotemark[$\dag$] [\timeform{''}]&Sgr A$^\ast$ in 2021\footnotemark[$*$] [\timeform{''}]&[\timeform{''}]&Sgr A$^\ast$ in 2017\footnotemark[$\dag$] [\timeform{''}]&Sgr A$^\ast$ in 2021\footnotemark[$*$][\timeform{''}]&[\timeform{''}]&[km s$^{-1}$]&[\timeform{D}]\\ 
\hline 
1&$-33.1038\pm0.0012$&$-33.0980\pm0.0014$&$-0.0058\pm0.0018$&$-17.9267\pm0.0013$&$-17.9268\pm0.0006$&$-0.0001\pm0.0013$&$57.4\pm22.7$&$91.0\pm13.7$\\
2&$-26.8390\pm0.0009$&$-26.8469\pm0.0016$&$0.0079\pm0.0018$&$23.6056\pm0.0009$&$23.6118\pm0.0007$&$0.0062\pm0.0012$&$99.4\pm21.8$&$-51.9\pm8.4$\\
3&$-26.4186\pm0.0015$&$-26.4093	\pm0.0018$&$-0.0093\pm0.0023$&$-34.4749\pm0.0017$&$-34.4833\pm0.0007$&$-0.0084\pm0.0018$&$124.3\pm29.2$&$132.3\pm	9.4$\\
4&$-23.0239\pm0.0065$&-&-&$-21.3101\pm0.0054$&-&-&-&-\\
5&$-22.7455\pm0.0031$&$-22.7323\pm0.0068$&$-0.0132\pm0.0074$&$16.9382\pm0.0035$&$16.9106\pm0.0033$&$-0.0276\pm0.0048$&$303.4\pm87.8$&$154.4\pm13.1$\\
6&$-19.9376\pm0.0007$&$-19.9323\pm0.0017$&$-0.0053\pm0.0019$&$33.6798\pm0.0007$&$33.6892\pm0.0009$&$0.0094\pm0.0011$&$106.8\pm21.5$&$29.7\pm9.1$\\
7&$-19.6078\pm0.0026$&	-&-&$30.0486\pm0.0016$&-&-&-&-\\	
8&$-18.6996\pm0.0008$&$-18.7185\pm0.0022$&$0.0189\pm0.0024$&$-26.0970\pm0.0009$&$-26.0971\pm0.0008$&$-0.0001\pm0.0012$&$186.8\pm26.4$&$-90.2\pm3.6$\\
9&$-13.8621\pm0.0011$&$-13.8630\pm0.0018$&$0.0008\pm0.0021$&$20.3542\pm0.0011$&$20.3498\pm0.0007$&$-0.0044\pm0.0013$&$44.0\pm24.6$&$-169.1\pm27.1$\\
10&$-12.4689\pm0.0005$&$-12.4747\pm0.0013$&$0.0058\pm0.0014$&$-11.0669\pm0.0005$&$-11.0605\pm0.0005$&$0.0064\pm0.0007$&$85.6\pm15.3$&$-42.2\pm7.4$\\
11&$-9.4867\pm0.0024$&$-9.4942\pm0.0028$&$0.0075\pm0.0037$&$11.8433\pm0.0017$&$11.8467\pm0.0013$&$0.0034\pm0.0022$&$81.7\pm42.8$&$-65.4\pm17.4$\\
12&$-7.6255\pm0.0004$&$-7.6119\pm0.0005$&$-0.0136\pm0.0006$&$-28.4701\pm0.0004$&$-28.4711\pm0.0002$&$-0.0011\pm0.0004$&$135.1\pm7.0$&$94.5\pm1.7$\\
13&$-3.7129\pm0.0017$&$-3.7216\pm0.0042$&$0.0087\pm0.0045$&$-4.7594\pm0.0015$&$-4.7807\pm0.0016$&$-0.0213\pm0.0022$&$228.2\pm49.5$&$-157.9\pm10.6$\\
14&$-3.6664\pm0.0014$&$-3.6551\pm0.0016$&$-0.0112\pm0.0021$&$-27.2653\pm0.0008$&$-27.2558\pm0.0005$&$0.0095\pm0.0010$&$146.2\pm22.9$&$49.7\pm6.0$\\	
15&$-3.2753\pm0.0003$&$-3.2796\pm0.0005$&$0.0043\pm0.0006$&$-6.9414\pm0.0003$&$-6.9514\pm0.0002$&$-0.0100\pm0.0003$&$107.9\pm6.9$&$-156.7\pm3.0$\\
16&$-1.1157\pm0.0009$&$-1.1021\pm0.0009$&$-0.0136\pm0.0013$&$-42.5625\pm0.0009$&$-42.5485\pm0.0003$&$0.0140\pm0.0010$&$193.7\pm16.0$&$44.0\pm3.4$\\
17&$-0.9775\pm0.0010$&$-0.9637\pm0.0021$&$-0.0139\pm0.0023$&$36.7772\pm0.0007$&$36.7710\pm0.0013$&$-0.00618\pm0.002$&$150.4\pm27.4$&$114.0\pm6.3$\\
18&$0.0330\pm0.0004$&$0.0328\pm0.0013$&$0.0002\pm0.0013$&$5.4697\pm0.0004$&$5.4535\pm0.0005$&$-0.0162\pm0.0006$&$160.9\pm14.4$&$-179.3\pm4.7$\\
19&$0.9450\pm0.0017$&$0.9657\pm0.0020$&$-0.0206\pm0.0026$&$-8.1656\pm0.0016$&$-8.1789\pm0.0008$&$-0.0132\pm0.0018$&$242.5\pm31.3$&$122.7\pm4.8$\\
20&$1.1890\pm0.0007$&$1.1836\pm0.0012$&$0.0054\pm0.00139$&$11.2088\pm0.0007$&$11.1855\pm0.0004$&$-0.0232\pm0.0008$&$236.5\pm15.8$&$-166.8\pm3.3$\\
21&$1.7635\pm0.0027$&$1.7748\pm0.0014$&$-0.0113\pm0.0030$&$40.3179\pm0.0025$&$40.3210\pm0.0005$&$0.00306\pm0.0025$&$116.4\pm39.3$&$74.87\pm12.6$\\
22&$2.3159\pm0.0017$&$2.3277\pm0.0013$&$-0.0118\pm0.0021$&$-15.0166\pm0.0015$&$-15.0246\pm0.0012$&$-0.00799\pm0.0019$&$141.0\pm28.5$&$124.2\pm8.0$\\
23&$5.7095\pm0.0005$&$5.7211\pm0.0009$&$-0.0116\pm0.0011$&$-6.3079\pm0.0005$&$-6.2947\pm0.0004$&$0.0132\pm0.0006$&$174.5\pm12.0$&$41.3\pm2.9$\\
24&$7.6841\pm0.0006$&$7.6850\pm0.0004$&$-0.0009\pm0.0007$&$4.1764\pm0.0006$&$4.1692\pm0.0001$&$-0.00727\pm0.0006$&$72.6\pm8.8$&$173.1\pm5.4$\\
25&$8.0885\pm0.0004$&$8.1028\pm0.0010$&$-0.0143\pm0.0011$&$-27.6538\pm0.0004$&$-27.6434\pm0.0004$&$0.0104\pm0.0006$&$175.4\pm12.3$&$54.0\pm2.6$\\
26&$9.3402\pm0.0012$&$9.3578\pm0.0008$&$-0.0176\pm0.0014$&$-18.3276\pm0.0010$&$-18.3266\pm0.0003$&$0.0010\pm0.0010$&$174.8\pm17.4$&$86.7\pm3.3$\\
27&$10.4954\pm0.0008$&$10.5020\pm0.0010$&$-0.0065\pm0.0013$&$-5.8878\pm0.0007$&$-5.9099\pm0.0003$&$-0.0221\pm0.0008$&$228.2\pm15.0$&$163.5\pm3.1$\\
 \hline
\end{tabular}}\label{tab:first}
\begin{tabnote}
\footnotemark[$\dag$] The position of Sgr A$^\ast$ was measured to be (RA, Dec)$_{\rm ICRS}$ = (\timeform{17h45m40.033456 \pm0.000002s}, \timeform{-29D00'28.22993 \pm0.00002''}) at September 19, 2017. \\
\footnotemark[$*$] The position of Sgr A$^\ast$  was measured to be  (RA, Dec)$_{\rm ICRS}$ = (\timeform{17h45m40.032847 \pm0.000005s}, \timeform{-29D00'28.26277 \pm0.00003''}) at August 19, 2021. \\
\footnotemark[$\ddag$] The angular differences between their positions relative to Sgr A$^\ast$ in 2017 and 2021 in Cartesian coordinate system.\\
\footnotemark[$\bot$] The Galactic center distance is assumed to be  $d =8.2$ kpc. The time difference between the observation epochs is $\Delta t = 3.917$ years.
\end{tabnote}
%\end{table}
\end{minipage}}\end{table*}
\normalsize
%%%%%%%%%%%%%%%%%%%%%%%%%%%%%%%%%%%%%%% %%%%%%%%%%%%%%%%%%%%%%%%%%%%%%
 \addtocounter{table}{-1}
%\clearpage
\small\begin{table*}[p]
\centering
\rotatebox{90}{\begin{minipage}{\textwidth} 
%\begin{table}
  \tbl{Continued. }{%
 \begin{tabular}{ccccccccc}
%\begin{tabular}{rrrrrrrrr}
  \hline
  No&$\Delta$RA relative to&$\Delta$RA realtive to&$\Delta X$\footnotemark[$\ddag$]&$\Delta$Dec relative to&$\Delta$Dec relative to&$\Delta Y$\footnotemark[$\ddag$]&$|V_{\mathrm{prop}}|$\footnotemark[$\bot$]&$PA$\\ 
&Sgr A$^\ast$ in 2017\footnotemark[$\dag$] [\timeform{''}]&Sgr A$^\ast$ in 2021\footnotemark[$*$] [\timeform{''}]&[\timeform{''}]&Sgr A$^\ast$ in 2017\footnotemark[$\dag$] [\timeform{''}]&Sgr A$^\ast$ in 2021\footnotemark[$*$][\timeform{''}]&[\timeform{''}]&[km s$^{-1}$]&[\timeform{D}]\\ 
\hline 
28&$13.1296\pm0.0005$&$13.1266\pm0.0006$&$0.0031\pm0.0007$&$5.5478\pm0.0005$&$5.5453\pm0.0002$&$-0.0025\pm0.0005$&$39.20\pm8.99$&$-129.4	\pm9.0$\\
29&$14.5733\pm0.0013$&$14.5805\pm0.0011$&$-0.0072\pm0.0017$&$-18.4720\pm0.0012$&$-18.4769\pm0.0004$&$-0.0049\pm0.0013$&$86.42\pm21.1$&$124.3\pm9.4$\\
30&$16.2562\pm0.0011$&$16.2667\pm0.0010$&$-0.0105\pm0.0015$&$-21.6657\pm0.0011$&$-21.6582\pm0.0003$&$0.0076\pm0.0012$&$128.3\pm18.7$&$54.3\pm5.7$\\
31&$17.2043\pm0.0012$&$17.2097\pm0.0014$&$-0.0054\pm0.0019$&$-4.8234\pm0.0011$&$-4.8361\pm0.0005$&$-0.0126\pm0.0013$&$136.2\pm22.3$&$156.7\pm7.4$\\
32&$22.1459\pm0.0030$&$22.1347\pm0.0046$&$0.0112\pm0.0055$&$15.3002\pm0.0046$&$15.2873\pm0.0029$&$-0.0129\pm0.0054$&$169.1\pm76.3$&$-139.1\pm18.3$\\
33&$22.5010\pm0.0011$&$22.4898\pm0.0024$&$0.0112\pm0.0026$&$23.4797\pm0.0011$&$23.4695\pm0.0008$&$-0.0102\pm0.0014$&$149.8	\pm29.3$&$-132.3\pm7.7$\\
34&$35.2757\pm0.0011$&$35.2951\pm0.0012$&$-0.0194\pm0.0016$&$30.7039\pm0.0011$&$30.7102\pm0.0004$&$0.0063\pm0.0011$&$202.0\pm19.3$&$72.0\pm3.3$\\
35&$39.0929\pm0.0024$&$39.1389\pm0.0089$&$-0.0460\pm0.0092$&$16.2455\pm0.0017$&$16.2314\pm0.0008$&$-0.0141\pm0.0019$&$477.2\pm93.0$&$107.1\pm3.9$\\
36&$40.9084\pm0.0009$&$40.9224\pm0.0004$&$-0.0141\pm0.0010$&$-22.0460\pm0.0011$&$-22.0426\pm0.0001$&$0.0034\pm0.0011$&$143.3\pm14.7$&$76.6\pm4.4$\\
37&$41.4085\pm0.0012$&$41.4129\pm0.0011$&$-0.0044\pm0.0016$&$15.1822\pm0.0011$&$15.1857\pm0.0004$&$0.0035\pm0.0012$&$55.87\pm19.6$&$52.0\pm13.7$\\
\hline
 \end{tabular}}\label{tab:first}
\begin{tabnote}
\footnotemark[$\dag$] The position of Sgr A$^\ast$ was measured to be (RA, Dec)$_{\rm ICRS}$ = (\timeform{17h45m40.033456 \pm0.000002s}, \timeform{-29D00'28.22993 \pm0.00002''}) at September 19, 2017. \\
\footnotemark[$*$] The position of Sgr A$^\ast$  was measured to be  (RA, Dec)$_{\rm ICRS}$ = (\timeform{17h45m40.032847 \pm0.000005s}, \timeform{-29D00'28.26277 \pm0.00003''}) at August 19, 2021. \\
\footnotemark[$\ddag$] The angular differences between their positions relative to Sgr A$^\ast$ in 2017 and 2021 in Cartesian coordinate system.\\
\footnotemark[$\bot$] The Galactic center distance is assumed to be  $d =8.2$ kpc. The time difference between the observation epochs is $\Delta t = 3.917$ years.
\end{tabnote}
%\end{table}
\end{minipage}}\end{table*}
\normalsize
%%%%%%%%%%%%%%%%%%%%%%%%%%%%%%%%%%%%%%

\appendix %%%%%%%%%%%%%%%%%%%%%%%%%%%%%%%%%%%%%%%%%%%%%%%%%%%%%%%%
%\onecolumn
 \section*{Line profiles of SiO $v=1, J=2-1$ Maser stars}
 Figures 10 show the line profiles of the SiO $v=1, J=2-1$ emission line toward the detected SiO maser stars except for 17454320-2900130. 
 \begin{figure} 
 \begin{center}
  \includegraphics[width=140mm]{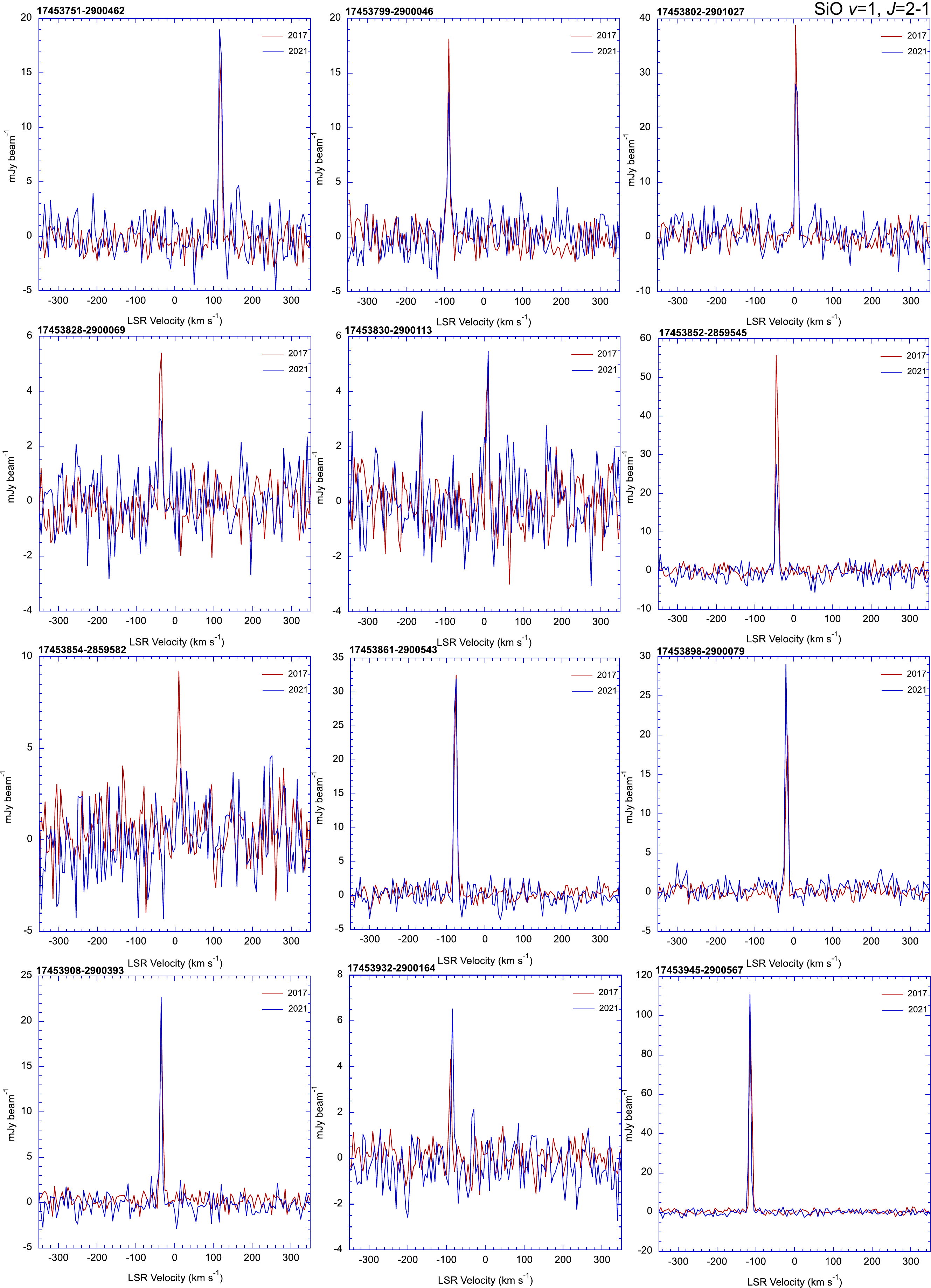}
 \end{center}
 \caption{Line profiles of SiO $v=1, J=2-1$ maser stars. The velocity interval  is 5 km s$^{-1}$.   {Alt text: 36 divisions. Two line graphs showing the line profiles at 2017 and 2021 toward a SiO maser stars in a division.}}
 \label{10}
\end{figure}
%%%
\addtocounter{figure}{-1}
\begin{figure}
 \begin{center}
  \includegraphics[width=140mm]{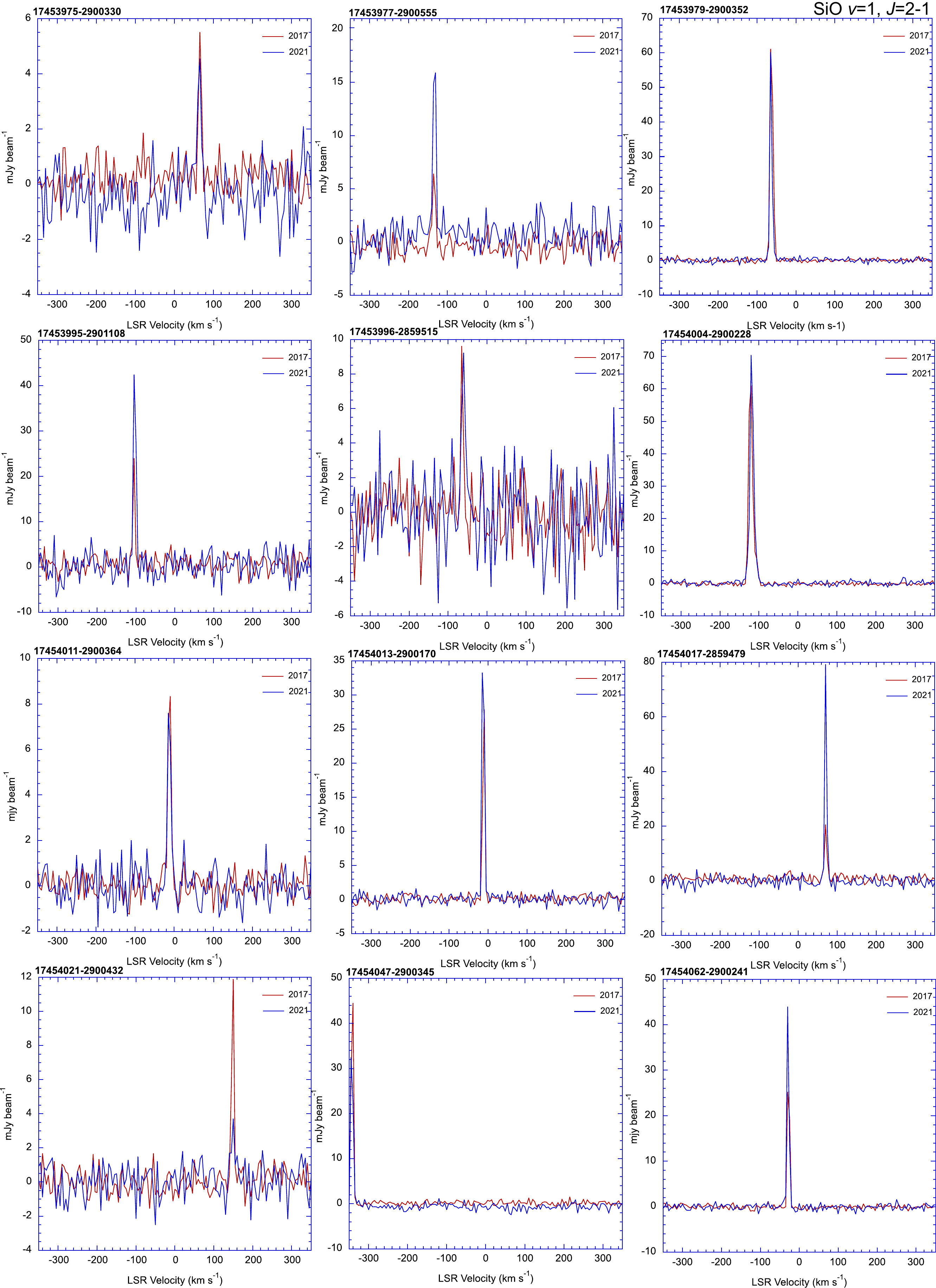}
 \end{center}
 \caption{Continued.}\label{.....}
\end{figure}
%%%
\addtocounter{figure}{-1}
\begin{figure}
 \begin{center}
  \includegraphics[width=140mm]{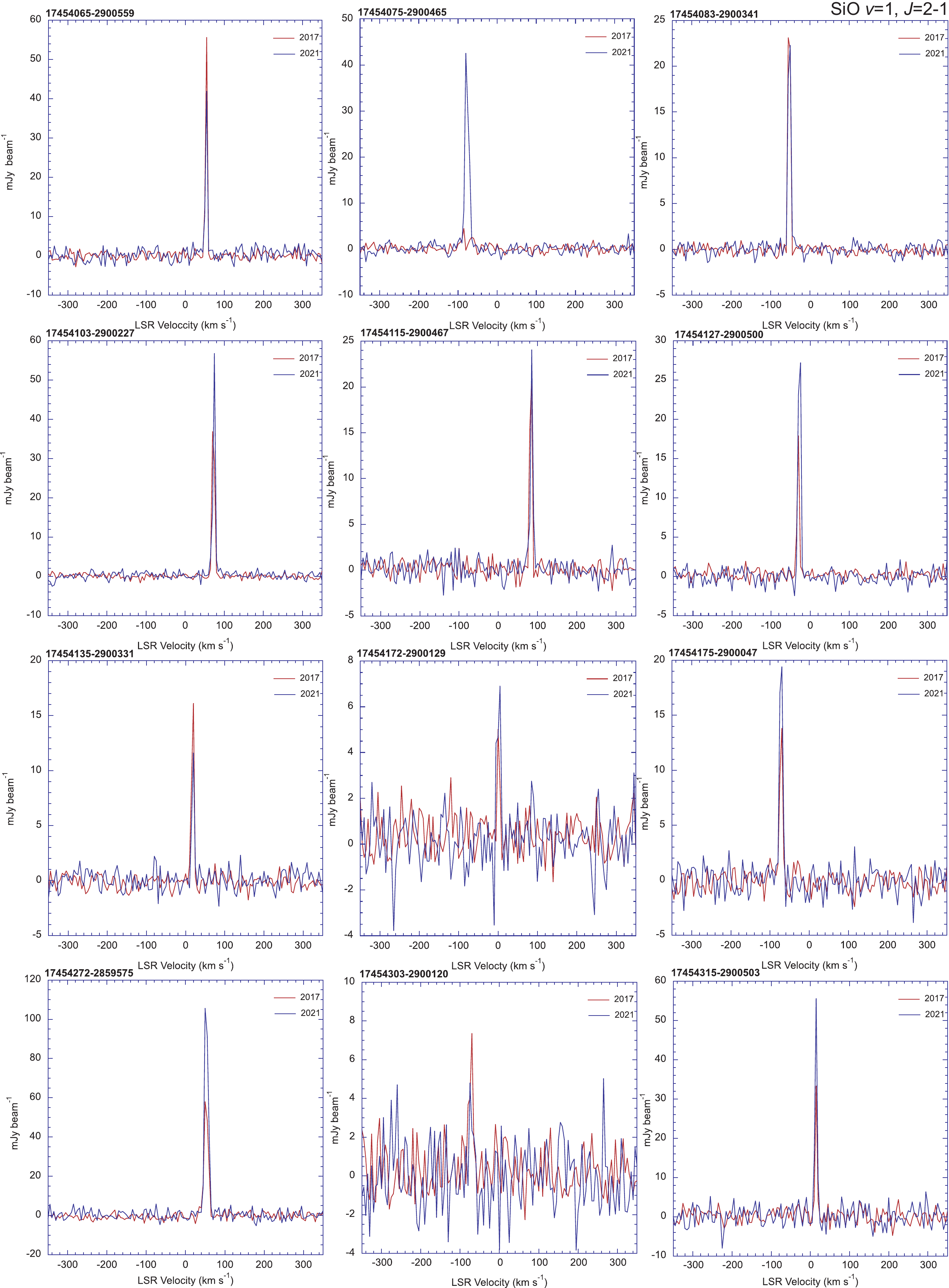}
 \end{center}
  \caption{Continued. }\label{.....}
\end{figure}

% Any journal's BST file (e.g., apj.bst) can be used as PASJ's BST is unavailable.    
% \bibliographystyle{****}
% \bibliography{****}
\twocolumn

\end{document}